# Electron-Spin Excitation Coupling in an Electron Doped Copper Oxide Superconductor

Jun Zhao[1,*,#], F. C. Niestemski[2,*,§], Shankar Kunwar[2], Shiliang Li[1,3], P. Steffens[4], A. Hiess[4], H. J. Kang[5], Stephen D. Wilson[2], Ziqiang Wang[2], Pengcheng Dai[1,6,3], & V. Madhavan[2]

[1]Department of Physics and Astronomy, The University of Tennessee, Knoxville, Tennessee 37996-1200, USA

[2]Department of Physics, Boston College, Chestnut Hill, Massachusetts 02467, USA

[3]Beijing National Laboratory for Condensed Matter Physics, Institute of Physics, Chinese Academy of Sciences, Beijing 100080, China

[4]Institut Laue Langevin, 6 Rue Jules Horowitz BP 156, F-38042 Grenoble Cedex 9, France

[5]NIST Center for Neutron Research, National Institute of Standards and Technology, Gaithersburg, Maryland 20899-8562, USA

[6]Neutron Scattering Science Division, Oak Ridge National Laboratory, Oak Ridge, Tennessee 37831, USA

[*]These two authors made equal contributions to the present paper.

[#]Present address: Department of Physics, University of California, Berkeley, California 94720-7300, USA

[§]Present address: Department of Physics, Stanford University, Stanford, California 94305, USA

Correspondence and requests for materials should be addressed to P.D. (pdai@utk.edu ) or V. M. (madhavan@bc.edu).

**High-temperature (high-$T_c$) superconductivity in the copper oxides arises from electron or hole doping of their antiferromagnetic (AF) insulating parent compounds. The evolution of the AF phase with doping and its spatial coexistence with superconductivity are governed by the nature of charge and spin correlations and provide clues to the mechanism of high-$T_c$ superconductivity. Here we use a combined neutron scattering and scanning tunneling spectroscopy (STS) to study the $T_c$ evolution of electron-doped superconducting $Pr_{0.88}LaCe_{0.12}CuO_{4-\delta}$ obtained through the oxygen annealing process. We find that spin excitations detected by neutron scattering have two distinct modes that evolve with $T_c$ in a remarkably similar fashion to the electron tunneling modes in STS. These results demonstrate that antiferromagnetism and superconductivity compete locally and coexist spatially on nanometer length scales, and the dominant electron-boson coupling at low energies originates from the electron-spin excitations.**

High-temperature (high-$T_c$) superconductivity in the copper oxides arises from electron or hole doping of their antiferromagnetic (AF) insulating parent compounds[1,2]. The evolution of the AF phase with doping and its spatial coexistence with superconductivity are governed by the nature of charge and spin correlations and provide clues to the mechanism of high-$T_c$ superconductivity[3,4]. Electron correlation in hole- and electron-doped materials can be quite different[5]. While doped electrons reside in the Cu $3d$ orbitals with correlations involving primarily $d$-electrons[2], doped holes form Zhang-Rice singlets[1] moving in the background of Cu spins and are subject to stronger correlations[1-7]. In electron-doped copper oxides, there are bulk signatures of coexisting AF and superconducting phases[8-12]. These measurements, however, cannot distinguish nanoscale spatial coexistence from larger scale phase segregation. Here we report advances made by a combined neutron scattering and scanning tunneling spectroscopy (STS) studies on nominally identical electron-doped superconducting $Pr_{0.88}LaCe_{0.12}CuO_{4-\delta}$ (PLCCO) samples with different $T_c$'s obtained through the oxygen annealing process[13,14]. We find that spin excitations detected by neutron scattering[15,16] have two distinct modes that evolve with $T_c$ in a remarkably similar fashion to the electron tunneling modes[17] in STS. Spatial mapping of the modes shows nanoscale regions of coexisting AF and superconducting order in the lower $T_c$ samples. Since the annealing process is not expected to change lattice (phonon) properties[2,13,14,18], these results demonstrate that antiferromagnetism and superconductivity compete locally and coexist spatially on nanometer length scales, and the dominant electron-boson coupling at low energies originates from the electron-spin excitations[17] rather than electron-phonon interactions[19].

There is experimental and theoretical evidence suggesting that antiferromagnetism is a competing phase to superconductivity in electron and hole doped copper oxides[1-5]. In samples where antiferromagnetism and superconductivity coexist, the spatial configuration of these two phases can provide important information on the strength of electron correlations[3,5] and the extent to which antiferromagnetism contributes to electron pairing[1-5]. In hole-doped copper oxides, such as $La_{2-x}Ba_xCuO_4$ near $x = 1/8$, strong electron correlation effects force the doped charge carriers into metallic rivers or stripes that are distributed as inhomogeneous patterns inside the AF insulating background[6,7]. Here, hole-doping of the parent compound $La_2CuO_4$ is achieved by replacing the trivalent

La$^{3+}$ with the divalent Ba$^{2+}$ which also induces chemical disorder and affects lattice properties[6,7,18], thus making it difficult to disentangle disorder from the effect of hole-doping. Electron-doped materials are uniquely suited to studying the competition between AF and superconducting orders. Unlike their hole-doped counterparts, electron-doping is carried out by replacing the (Pr,La)$^{3+}$ in the parent compound (Pr,La)$_2$CuO$_4$ with the Ce$^{4+}$ to form a non-superconducting antiferromagnet Pr$_{1-x}$LaCe$_x$CuO$_4$ (refs. 2,8). Further post-growth annealing treatment is required to suppress the static AF order and obtain superconductivity[2]. Since the annealing process only removes a tiny amount of excess oxygen and has minimal effects on its lattice structure[2,13,14,18], electron-doped PLCCO provides an unique platform to study the evolution from AF order to superconductivity through the oxygen annealing process[10,11,13-16].

Here we report a combined neutron scattering and STS study on nominally identical electron-doped superconducting PLCCO samples with $T_c$'s of 21 K and 24 K obtained through the oxygen annealing process[14]. The $T_c$ = 24 K PLCCO system is a pure superconductor, while the $T_c$ = 21 K sample has static AF order coexisting with superconductivity[10,11,14]. Using polarized and unpolarized neutron triple-axis spectroscopy (see supplementary information), we show that spin excitations in the $T_c$ = 24 K PLCCO have two modes near 2 meV and 10.5 meV (ref. 15). Upon annealing to obtain the $T_c$ = 21 K PLCCO system, the intensity of the 2 meV mode increases dramatically while the 10.5 meV mode is downshifted to 9.5 meV and becomes much weaker across $T_c$ (refs. 11,14). Remarkably, our STS measurements on the same samples also reveal two modes that evolve with $T_c$ in an almost identical manner. A comparison of the spatial and temperature dependence of the neutron and STS modes suggests that the 2 meV mode is associated with antiferromagnetism while the ~10 meV mode is connected with superconductivity. Since the oxygen annealing process that changes $T_c$ from 21 K to 24 K in PLCCO is not expected to affect phonons, our data indicate that spin excitations are indeed observed in STS data. If high-$T_c$ superconductivity in copper oxides requires a bosonic "pairing glue" [20-22], these results would suggest that spin excitations are the mediating glue for the electron pairing and superconductivity in PLCCO.

In previous unpolarized neutron scattering experiments on the $T_c$ = 24 K PLCCO, a resonance mode centered at the AF wavevector $Q$ = (1/2, 1/2) was observed near 10.5 meV (ref. 15). To confirm the magnetic nature of the resonance, we carried out neutron polarization analysis on the $T_c$ = 24 K sample (Fig. 1a). For neutron polarizations along the $Q$-direction, magnetic scattering flips the polarization direction of the incident neutrons (neutron spin-flip or SF) while neutron non-spin-flip (NSF) scattering probes pure nuclear scattering[23]. The neutron polarization analysis can therefore unambiguously separate magnetic scattering from nonmagnetic scattering processes[23]. Similar $Q$-scans were obtained for both the SF and NSF scattering at 2 K with energy transfers near the resonance ($E$ = 10 meV) and well below it ($E$ = 3 meV) (Figures 1b-1e). While a peak centered at the AF ordering wavevector is observed for the SF channel (Figs. 1b,1c), the NSF scattering is featureless (Figs. 1d,1e). These data indicate that the excitations at 10 meV and 3 meV near $Q$ = (1/2,1/2) are entirely magnetic in origin without any lattice contribution to the scattering. On warming to 30 K (Fig. 1f), the intensity of the 10 meV excitations decrease, and are consistent with a prototypical neutron spin resonance mode[15]. Figure 1g shows that SF scattering for neutron polarization direction parallel to the $Q$-direction is about twice as large as those for perpendicular to it. This indicates that the resonance is due to isotropic paramagnetic scattering[23] and is consistent with the mode being a singlet-to-triplet excitation associated with electron pairing[24].

To determine the energy dependence of the dynamic spin susceptibility $\chi''(Q,\omega)$, we measure energy scans at the AF wavevector $Q = (1.5,-0.5)$ and background $(1.64,-0.36)$ positions above and below $T_c$. Figures 2a and 2b show the raw data for SF and NSF scattering, respectively. The SF magnetic data in Fig. 2a are dominated by the wavevector independent $Pr^{3+}$ crystalline electric field (CEF) level near 18 meV (refs. 16,25), while the weak peak near 16 meV in the NSF scattering is due to imperfect neutron polarization and nonmagnetic phonon scattering. Figure 2c shows $\chi''(Q,\omega)$ at $Q = (1.5,-0.5)$ in absolute units, obtained by subtracting the background scattering including $Pr^{3+}$ CEF, including low-energy data (see supplementary information), correcting for the Bose population factor, and normalizing to acoustic phonons[26]. The outcome reveals two broad peaks at ~2 meV and 10.5 meV that are enhanced below $T_c$. Solid lines in this plot are simple guides to the eye fit by two Gaussians to identify the mode energies. To determine the $T_c$ evolution of the resonance and 2 meV excitations, we plot in Figure 2d $\chi''(Q,\omega)$ for both the $T_c = 24$ K and 21 K PLCCO (refs. 11,14,16, and supplementary information). We find that the effect of decreasing $T_c$ from the 24 K to 21 K in PLCCO through the annealing process is to enhance the 2 meV mode and reduce the intensity of the neutron spin resonance near 9 meV (Fig. 2d).

We now turn to STM measurements that can directly probe the superconducting state and its spatial distribution. PLCCO samples were cleaved in UHV and directly inserted into the STM head held at 5 K. Data from five samples and tips (three 24 K PLCCO samples and two 21 K PLCCO samples) are included in this paper. We studied multiple spots within each sample, which also contributes to our robust set of statistics. Due to lack of a periodic lattice on the surface of PLCCO, other methods of tip and sample characterization are crucial for ensuring the reliability of the data. Tips were characterized on the prototypical superconductor $Bi_2Sr_2CaCu_2O_8$. Scanning electron microscopy (SEM) studies reveal large micron sized flat regions, which were accessed using our ability to move to different regions with the coarse walking mechanism of the STM (more details on tip and sample characterization are described in the supplementary information).

Compared with the $T_c = 24$ K PLCCO, the 21 K samples have smaller superconducting gaps ($5.5 \pm 0.5$ meV versus $7.7 \pm 1.2$ meV for the 24 K samples, Fig. 3a) that disappear just above the bulk $T_c$ and below $T_N$ (see Fig. SI4). We note that the smaller value of the standard deviation for the statistical averages for the 21 K sample are possibly as a result of a smaller statistical pool for the 21 K PLCCO compared to the 24 K PLCCO, which makes the distribution more narrow. It is, however, clear that the entire gap histogram is shifted to lower energies for the 21 K sample resulting in a lower average gap. In general, the 21 K gaps have muted coherence peaks (Figs. 3b and 4d) compared to the 24 K samples[17]. The height and width of the coherence peak are a measure of the scattering rate ($\Gamma$), which contains a temperature dependent contribution from phonons and electronic excitations as well as a temperature independent contribution from impurity scattering ($\Gamma_{imp}$) (refs. 27, 28). The smaller height (larger width) of the coherence peaks at the same measurement temperatures as 24 K samples indicate a larger impurity $\Gamma_{imp}$ in the 21 K PLCCO. Since the two samples have nominally identical Ce-concentrations, the additional disorder is likely due to the in plane defects, which is consistent with more in-plane Cu vacancies in the lower $T_c$ sample[13].

In the 21 K samples, outside the gap ($\Delta$) we observe three distinct satellite features labeled $E_A$, $E_B$ and $E_C$ in Fig. 3b, that originate from the coupling of electrons to collective excitations (bosonic modes, which can be either spin excitations[17] or phonons[29]) in the sample. Our first task is to obtain the energy of these bosonic modes from the STS spectra. In a superconductor, a bosonic excitation at energy $\Omega$ results in a feature at energy ($E$) offset by the gap in the tunneling spectroscopy[17,29]. The local bosonic mode energy can therefore be determined from the STS spectrum by subtracting the local gap energy scale, i.e., $\Omega = E - \Delta$. The energy $E$ is best determined (see supplementary information) by locating the position of the maxima/minima in the second derivative of the tunnel current, $d^2I/dV^2$, above and below the Fermi energy respectively (Fig. 3b). We first consider the higher energy features, labeled $E_B$ and $E_C$ in Fig. 3b.

Statistics for $\Omega_B$ and $\Omega_C$ (Fig. 3d) reveal that the average energies are 9.2 ± 0.85 meV (we henceforth term this mode $\Omega_R$) and 19.2 ± 2.1 meV. Standard deviations were calculated from the Gaussian fit to the histograms, as shown in 3d. These energies are reminiscent of the modes found in our earlier STS studies of 24 K PLCCO samples (~10.5 meV and 21 meV)[17] where the neutron spin resonance is found to occur at ~11 meV (ref. 15, Fig. 2c). The correspondence in energy combined with the lack of a peak in the phonon density of states at ~11 meV (Fig. 2b) had suggested a common origin for the STS and neutron modes with the 21 meV feature possibly arising from a second harmonic process[17]. We note that there are phonon modes at energies above 15 meV[17], which could also contribute to the spectral intensity in STM spectra. This may explain why in some cases the intensity of $\Omega_C$ is comparable to that of $\Omega_B$. STM data cannot distinguish between these possibilities for the higher energy mode; however our unique combined STM and neutron studies suggest a magnetic origin. Most importantly, this does not affect any conclusions about $\Omega_B$. In the 21 K PLCCO, the same neutron spin resonance is found to shift to a lower energy of ~ 9.3 meV in accordance with the lower $T_c$ (ref. 14, Fig. 2c). The corresponding downward energy shift of the STS mode (Figs. 3c and 3d) provides strong support for its identification with the spin-resonance. Consistent with this picture, the broad higher energy feature is also shifted to lower energies. We note that similar to the 24 K sample[17], the intensity of the spin resonance mode in the 21 K data shows spatial variations. But while the mode is observed in ~85% of spectra in the 24 K samples, the 21 K samples show more areas where the intensity of the mode is suppressed such that it is picked up only in ~63% of the spectra. This is consistent with the fact that the 21 K sample has weaker superconducting heat capacity anomaly[14].

Having identified the 9.2 meV mode with the neutron spin-resonance, we turn to the lower energy feature, labeled $E_A$ in Fig. 3b. From the mode statistics (Fig. 3d), we obtain an average energy of $3.0 \pm 0.3$ meV and term this new mode $\Omega_{AF}$. Since there is no peak in the phonon density of states at these energies (Fig. 2b), we look at electronic excitations to explain its origin. As-grown PLCCO exhibits three-dimensional AF order below $T_N \approx 200$ K (ref. 10). Neutron scattering data show that upon annealing to remove oxygen and obtain the 21 K sample, $T_N$ reduces to ~40 K and there is a drastic renormalization of the spin dynamics with a peak in the low temperature local spin susceptibility ($\chi''(\omega)$) appearing near ~2 meV (refs. 10,11,16). This low energy mode is reminiscent of the overdamped excitation observed to emerge in hole-doped $YBa_2Cu_3O_{6.35}$ close to the AF phase boundary[30] and is a potential candidate for $\Omega_{AF}$. In the neutron data, the 2 meV neutron peak in the 21 K sample is rapidly suppressed in the 24 K (refs. 11,15,16). This is in stark contrast to the spin-resonance mode (~10 meV) whose energy and spectral weight are enhanced upon tuning toward higher $T_c$'s (Fig. 2c). In remarkable agreement with the neutron scattering data, we find that $\Omega_{AF}$ is more prominent in the 21 K sample compared to the 24 K sample where this mode is only rarely observed; and quite opposite to the behavior of $\Omega_R$. The correspondence in energy and doping dependence make it highly credible that $\Omega_{AF}$ and $\Omega_R$ have a magnetic origin similar to the neutron modes with the same energy.

Keeping in mind that $\Omega_{AF}$ is a signature of local AF order, we construct spatial maps for $\Omega_{AF}$ and $\Omega_R$ (Figs. 4b and c) in the 21 K sample. Comparing the maps we find that there are nanometer-sized regions where $\Omega_{AF}$ and $\Omega_R$ are both present. These areas are not macroscopically segregated but are rather fully embedded within the superconducting regions. The data clearly show that AF order and superconductivity coexist at nanometer length scales. The question remains whether the coexistence is ubiquitous or confined to nanometer-sized patches. From the neutron studies, we know that the 21 K sample has macroscopic AF phase coherence associated with static long-range AF order[10]. Therefore, it seems likely that $\Omega_{AF}$ should also exist in extended, contiguous regions of the sample. There are two possible explanations for the seemingly contrary nature of $\Omega_{AF}$ in the STS data. Firstly, similar to $\Omega_R$ the varying intensity of the mode makes it difficult to discern in some regions. However, there is a second, more compelling explanation. As seen in the histogram (Fig. 3d), the mode energy varies reflecting the spatial inhomogeneity of the AF order. Since the mode lies close to the gap edge, a small decrease in the mode energy pushes it toward the gap edge, merging it with the coherence peak. This is clearly seen in the line-cut (Fig. 4d) as we go from a coexisting region (spectrum 1) to the area where the mode is indistinguishable from the coherence peak (spectrum 36). A closer analysis of the STS data thus indicates that the mode might well be widespread but simply difficult to see when it merges with the coherence peak implying that the STM analysis likely overestimates the energy of $\Omega_{AF}$. Since the neutron data represent an average over all regions, this accounts for the smaller neutron mode energy.

This scenario is further supported by the gap map (Fig. 4a), which shows clear spatial correlation with the $\Omega_{AF}$ mode map. Comparing the two points to the competing nature of the two coexisting orders: regions with a prominent $\Omega_{AF}$ mode are found with smaller superconducting gaps (and minimal coherence peaks). The larger gap regions are associated with an underlying smaller and weaker $\Omega_{AF}$. We therefore conclude that the raw STS data most likely underestimates of the size of the coexisting region. This in conjunction with the neutron data, illustrates a picture of a long-range ordered AF phase in the 21 K compound which weakens on further annealing into short-ranged, local AF order and correspondingly higher $T_c$'s, as seen in the 24 K sample. Moreover, given that one does not expect phonons to change for the 21 and 24 K samples, our results (Figs. 2c and 3c, 3d) represent the most compelling evidence that spin excitations are intimately associated with tunneling electronic bosonic modes observed by STS.

**Acknowledgements** The neutron scattering work at UT/ORNL is supported by the U.S. NSF-OISE-0968226, and by the U.S. DOE, Division of Scientific User Facilities. Work at BC is supported by U.S. NSF and DOE. The single crystal PLCCO growth effort at UT is supported by U.S. DOE BES under Grant No. DE-FG02-05ER46202. Work at IOP is supported by the Chinese Academy of Sciences, the Ministry of Science and Technology of China (973 Project nos. 2010CB833102 and 2010CB923002). J.Z. is supported by a fellowship from Miller Institute of Basic Research in Science at Berkeley.

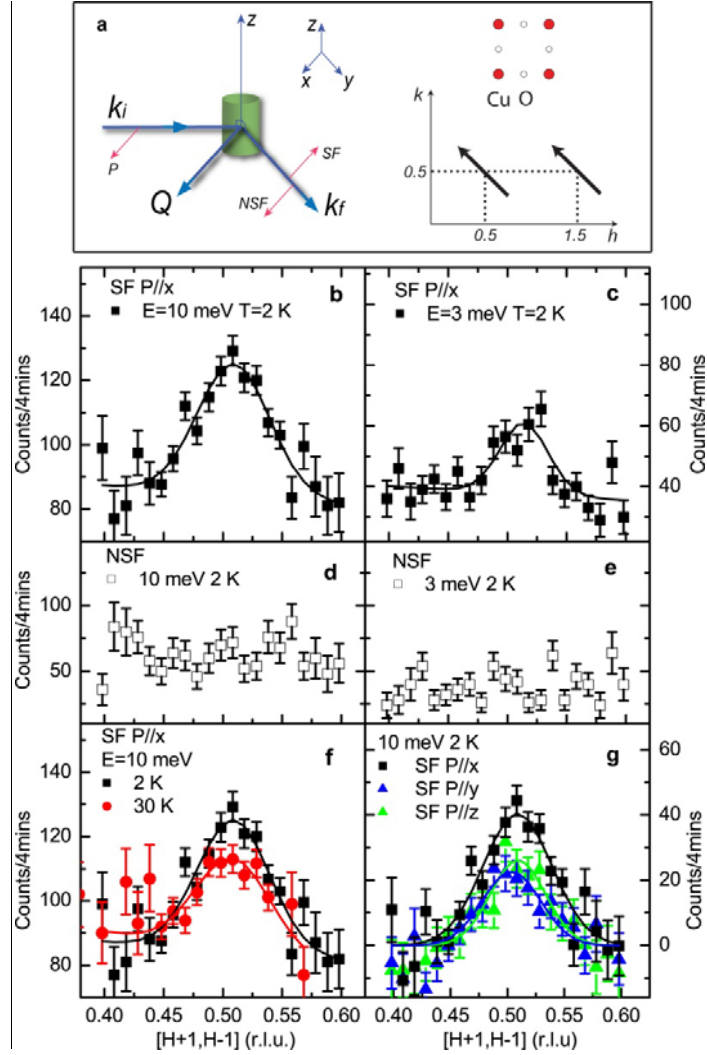

**Figure 1 Schematic diagram of polarized neutron scattering set up and polarized neutron scattering data at various temperatures.** (a) Real and reciprocal space of the $CuO_2$ plane. The position in reciprocal space at wavevector $\mathbf{Q} = (q_x, q_y, q_z)$ Å$^{-1}$ is labeled as $(H, K, L) = (q_x\, a/2\pi,\, q_y\, b/2\pi,\, q_z\, c/2\pi)$ reciprocal lattice units (rlu), where the tetragonal unit cell of PLCCO (space group $I_4/mmm$) has lattice parameters of $a = b = 3.98$ Å, $c = 12.27$ Å. Our polarized neutron scattering experiments were carried out on the IN20 thermal neutron three-axis spectrometer at Institut Laue Langevin with a fixed final neutron energy of $E_f = 14.7$ meV. With the Cryopad setup, we can study the magnetic excitations of PLCCO in a strictly zero magnetic field (< 10 mG), thus avoiding errors due to flux inclusion or field expulsion in the superconducting phase of the sample. (b) $\mathbf{Q}$-scans through $[\mathbf{Q} = (1.5, -0.5, 0)]$ at $E = 10$ meV in the SF channel at 2 K; (d) NSF channel; (c),(e) Identical $\mathbf{Q}$-scans at $E = 3$ meV and 2 K in the SF and NSF channels, respectively. (f) Temperature dependence of the SF scattering at $E = 10$ meV. (f) Neutron spin polarization dependence of the resonance at $E = 10$ meV. The solid lines are Gaussian fits on linear backgrounds.

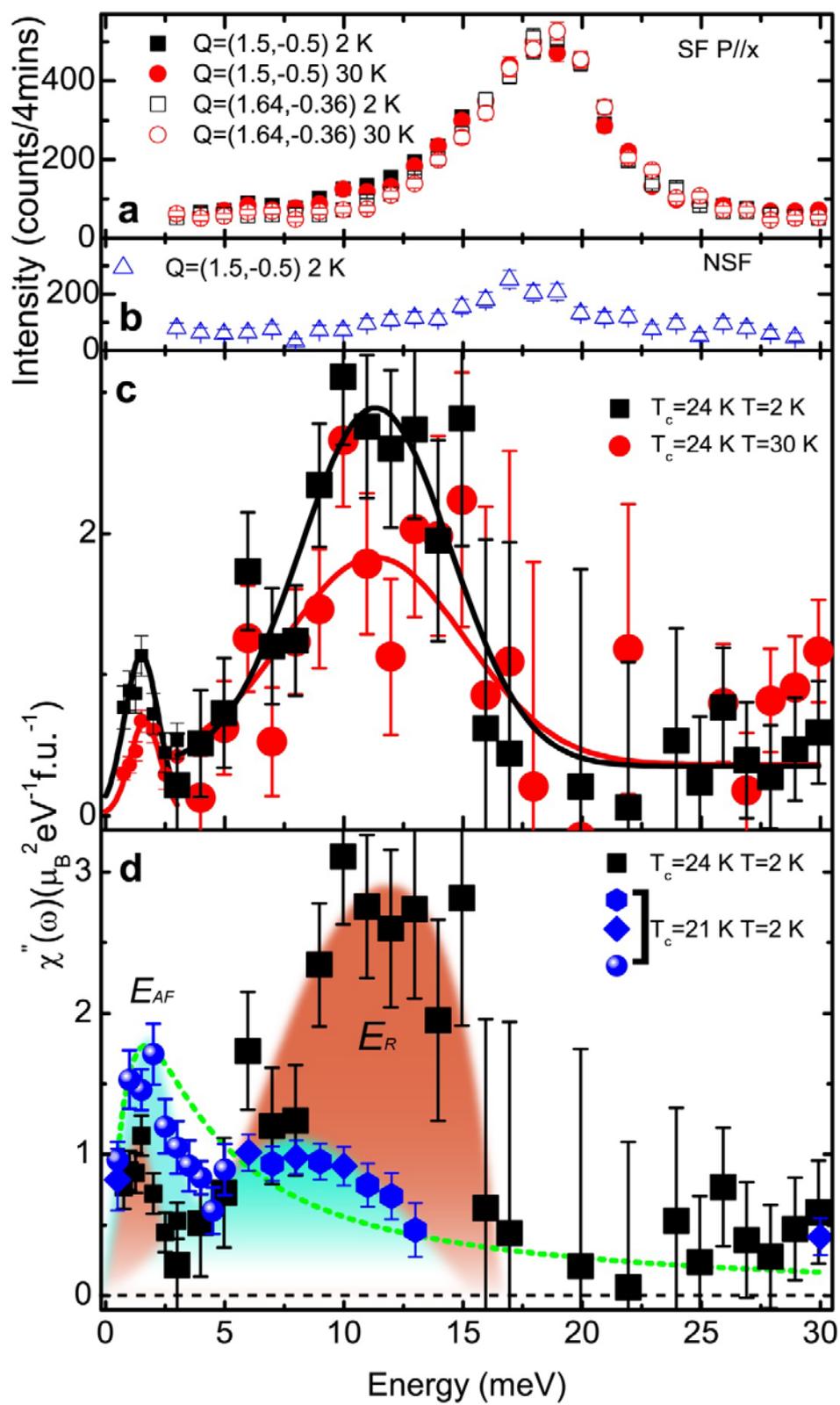

**Figure 2 Energy dependence of SF and NSF scattering at $Q = (1.5, -0.5, 0)$ rlu for the 24 K PLCCO and $\chi''(\omega)$ in absolute units for the 21 and 24 K samples.** (a) energy scans at the signal $Q = (1.5,-0.5,0)$ and background $Q = (1.64,-0.36,0)$ positions at 2 K and 30 K in the SF channel, the Q independent 18 meV peak originates from the CEF excitation of $Pr^{3+}$ (ref. 25); (b) Energy scans at $Q = (1.5,-0.5,0)$ in the NSF channel. (c) The combined and normalized low-energy $\chi''(\omega)$ above and below $T_c$ for the 24 K PLCCO in absolute units. The black and red solid lines are guides-to-the-eye based on weighted Gaussian fit for 2 K and 30 K data, respectively. (d) Comparison of the $\chi''(\omega)$ for the 24 K and 21 K PLCCO at 2 K. The data for the 21 K PLCCO are from Figs. 5(a) and 6 (a)-(c) of Ref. 11 (solid blue circles) normalized to the data in absolute unit from Ref. 16 (diamond shaped squares). The solid hexagons are data from Ref.14 normalized with data from Refs. 11 and 16. The green dashed line shows an attempted fit using a modified Lorentzian, which clearly cannot fit the data.

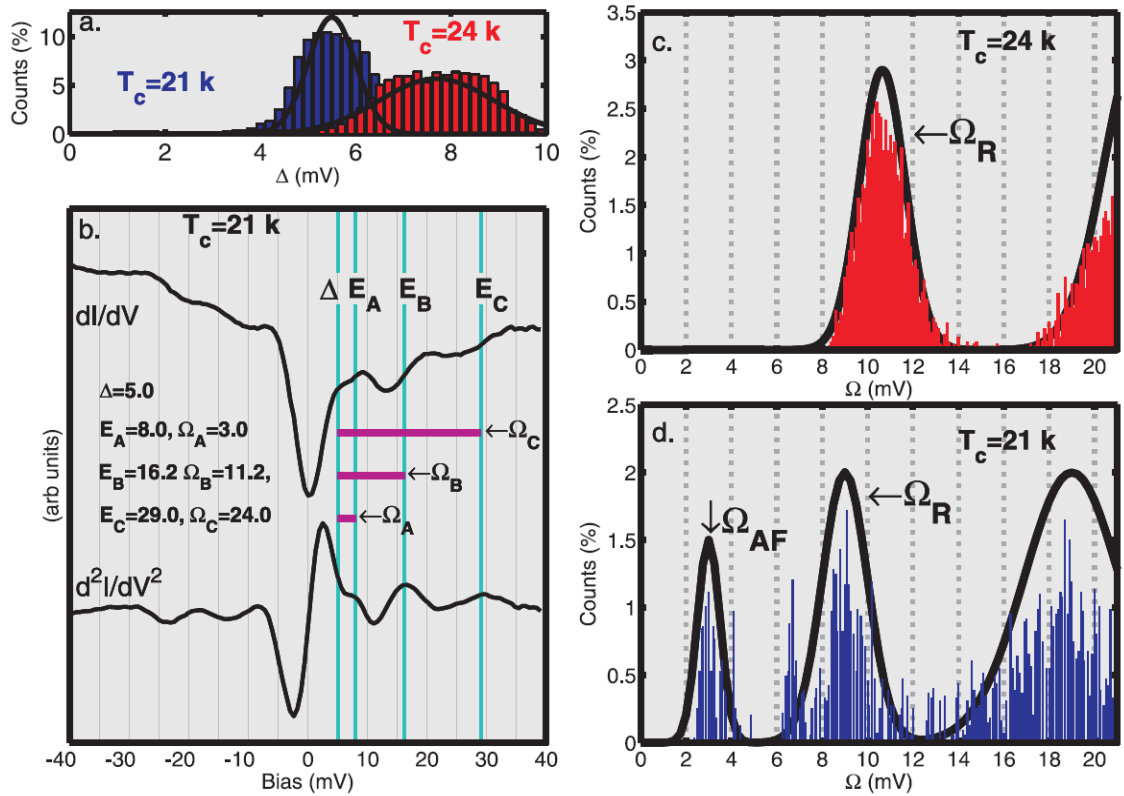

**Figure 3 Comparison of the tunneling spectra of the 21 and 24 K samples.** Our STS measurements were performed on the same instrument as described before[17]. (a) Histogram showing the distribution of the superconducting gaps for the 21 K and 24 K PLCCO samples. (b) Single *dI/dV* spectrum of the 21 K PLCCO and its derivative *(d²I/dV²)* offset below, demonstrating the existence of the superconducting gap and the three modes in the same spectrum. The derivative was smoothed (using nearest neighbor averaging) and multiplied by a simple factor. The spectrum was obtained with a junction resistance of 16.5 Mohms. The energies of the local gap ($\Delta$) and the three features ($E_A$, $E_B$, and $E_C$) from which the corresponding mode energies are calculated are shown on the figure. (c, d) Comparison of the mode statistics for the 21 K and 24 K samples. A Gaussian fit to the data is shown as a guide to the eye. The 21 K samples shows three modes, $\Omega_{AF} = 3.0 \pm 0.3$ meV, $\Omega_R = 9.2 \pm 0.85$ meV, and a third mode at $19.2 \pm 2.1$ meV. Note that $\Omega_A$ and $\Omega_B$ in b. are now termed $\Omega_{AF}$ and $\Omega_R$. The 24 K data shows the primary mode ~10.5 meV and part of the second mode ~21 meV. The 21 K $\Omega_R$ modes are shifted with respect to the 24 K sample. The 3 meV, $\Omega_{AF}$ mode, which is clearly visible in the 21 K data is rarely observed in the 24 K data and therefore does not have a presence in the histogram.

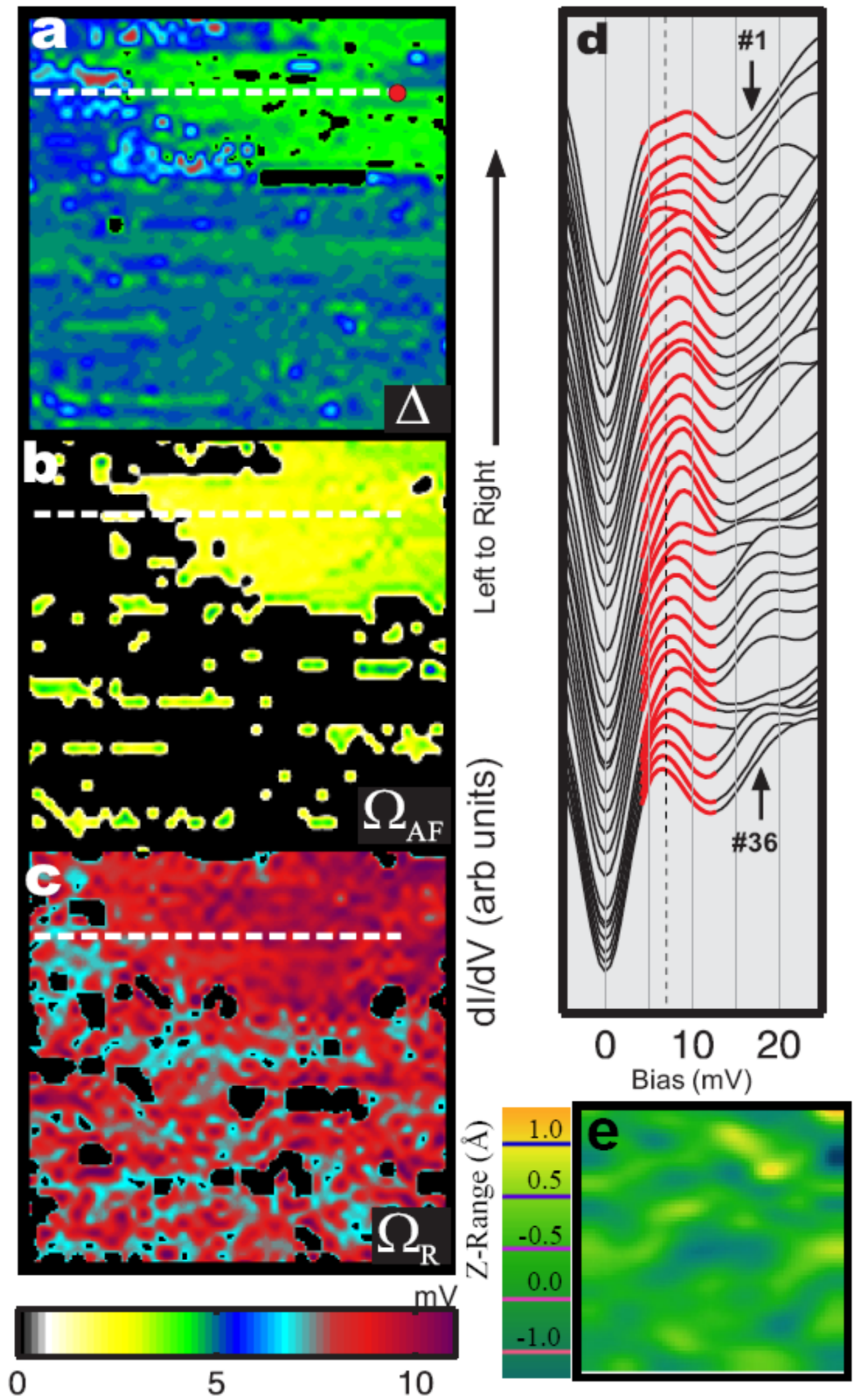

**Figure 4** **90 Å maps of the same area of the 21 K sample showing** (a) $\Delta$, (b) $\Omega_{AF}$, and (c) $\Omega_R$. The maps have the same color scale, which is shown at the bottom. The dashed line shows the position of the 70 Å linecut in (d). (d) The linecut shows the transition from a spectrum showing a clear low energy mode $\Omega_{AF}$ (labeled #1, shown as a red dot on the $\Delta$ map) to a spectrum (labeled #36) at the end of the linecut, where the coherence peak and $\Omega_{AF}$ seem to have merged into one entity. The dotted line traces the low energy mode as it evolves. The energy range of interest is highlighted in red. The spectra were obtained at 2 Å intervals and have been offset for clarity. All spectra were obtained with a junction resistance of 16.5 Mohms. (e) Topography in the same region as the maps.

# Supplementary Information (SI) for

# Electron-Spin Excitation Coupling in an Electron Doped Copper Oxide Superconductor

Jun Zhao, F. C. Niestemski, Shankar Kunwar, Shiliang Li, P. Steffens, A. Hiess, H. J. Kang, Stephen D. Wilson, Ziqiang Wang, Pengcheng Dai , & V. Madhavan

In addition to polarized neutron scattering measurements, we have carried out unpolarized neutron measurements on the 24 K PLCCO using cold neutron triple-axis spectroscopy on Spin-Polarized Inelastic Neutron-Scattering Spectrometer (SPINS) at NIST Center for Neutron Research, Gaithersburg, Maryland. The final neutron energy was fixed at $E_f$ = 5 meV with collimations guide-open-80'-open. A cooled Beryllium filter was put behind the sample to eliminate higher-order contamination of the scattered beam. Figures SI1a-1d summarize the low energy constant-$E$ scans along the [$H$,1-$H$,0] direction at different temperatures. Consistent with earlier results[1-4], the magnetic excitations are commensurate and centered at $Q$ = (0.5, 0.5, 0) for all energies probed. In previous work, our data suggest that magnetic excitations between 1.5 and 4.5 meV are weakly temperature dependent from 2 K to 55 K (ref. 1). The new measurements with much higher statistics at $E$ = 1 meV in Fig. SI1a show that the spin excitation intensity reduces on cooling from the normal state (24 K) to the superconducting state (2 K). In addition, the scattering at $Q$ = (0.5, 0.5,0) disappears at 150 K. For energies between 1.5 and 3.5 meV, the excitations are essentially temperature independent between 2 K and 24 K consistent with earlier work[1], but also vanish at 150 K (Fig. SI1b).

To accurately determine the energy dependence of the dynamic spin susceptibility $\chi''(Q,\omega)$, we carried out constant-$Q$ scans at the peak center (0.5,0.5,0) and background (0.56,0.44,0) positions at different temperatures. Figure SI2a shows the data at 2 K and 24 K. By subtracting the backgrounds and correcting for the Bose population factor, we obtain $\chi''(Q,\omega)$ at $Q = (0.5, 0.5, 0)$ at different temperatures as shown in Fig. SI2b. Inspection of the Figure reveals that $\chi''(Q,\omega)$ increases systematically on cooling and has a characteristic energy scale around 2 meV at 2 K. To model its temperature dependence, we fit the observed inelastic response to a modified Lorentzian with the relaxation rate $\Gamma$ via $\chi''(\omega) \propto \Gamma\omega/(\Gamma^2 + \omega^2)$, similar to previous work on low-energy spin excitations of $Pr_{1-x}LaCe_xCuO_{4+\delta}$ (refs. 4,5). The solid lines in Fig. SI2b show the best fits to the data, where the temperature dependence of the spin fluctuation relaxation rate $\Gamma$ is plotted in the inset with $\Gamma \propto (1.02 \pm 0.2)T$. This suggests that spin excitations above $T_c$ are relaxational with a single energy scale controlled by temperature, much like spin fluctuations in an AF Fermi-liquid[6]. At 2 K ($\ll T_c$), the modified Lorentzian form fails to describe the data above 4 meV, and the data can instead be described by a localized mode around 2 meV on a continuum of scattering. This is unexpected from the simple spin exciton model for the neutron spin resonance[7].

Figure SI3 shows the temperature dependence of the scattering at energies below and above the 2 meV energy scale. The peak intensity at 1 meV and $Q = (0.5, 0.5, 0)$ displays a clear suppression near $T_c$ suggesting the opening of a pseudo spin gap in $S(Q,\omega)$, much like that of underdoped $La_{1.895}Sr_{0.105}CuO_4$ (ref. 8) and $YBa_2Cu_3O_{6+y}$ (ref. 9). The background scattering at $Q = (0.56, 0.44, 0)$ shows monotonic increase with increasing temperature and no anomaly across $T_c$ (Fig. SI3a). Figures SI3c and 3e show

the temperature dependence of the magnetic scattering at 1 meV and the corresponding $\chi''(Q,\omega)$, at $Q = (0.5, 0.5, 0)$, respectively. It is clear that magnetic scattering disappears around 150 K, and $\chi''(Q,\omega)$ saturates approximately below $T_c$. For magnetic scattering at 3.5 meV, our data show very weak temperature dependence of the scattering consistent with earlier results (Fig. SI3b)[1]. The scattering above background shows no anomaly across $T_c$ (Fig. SI3d), and $\chi''(Q,\omega)$ increases gradually from 150 K with decreasing temperature (Fig. SI3f).

To determine the $\chi''(Q,\omega)$ in absolute units for the 21 K PLCCO, we normalize triple-axis data in Refs. 2,3 to time of flight data in absolute units (Ref. 4). Figures SI4a-c show $\chi''(Q,\omega)$ below and above $T_c$ from constant energy scans from Ref. 2. Inspection of the data reveals a clear peak at 1.5 meV in $\chi''(Q,\omega)$ at 2 K. Similarly, one also sees a clear superconductivity induced enhancement of $\chi''(Q,\omega)$ near 9 meV as shown in constant-energy scans (Fig. SI4d). The weak resonance is also confirmed in the temperature difference plot shown in Fig. SI4e (Ref. 3). To obtain the $\chi''(Q,\omega)$ in absolute units, we then compare constant-Q and constant-energy scans from Refs. 2 and 3, and normalize them to low energy data in Ref. 4. The outcome for the $\chi''(Q,\omega)$ is shown in Fig. SI5 which shows two modes near 2 meV and 9 meV. A comparison of the 21 K and 24 K PLCCO data in absolute units is shown in Fig. 3d.

For our STS measurements, low temperature (5.5 K) spectra on the 21 K samples reveal a gap at the Fermi energy (Fig. SI6). Since the gap could arise from either superconductivity or static AF order, temperature dependent spectra are essential to identify its origin. STS spectra obtained as a function of temperature show that the average gap disappears above $T_c$ but below $T_N$. Furthermore, consistent with a lower $T_c$, the average gap is smaller, (Fig. 3a) in the 21 K samples (5.5 ± 0.5 meV) compared to the 24 K samples (7.7 ± 1.2 meV). This gap can therefore be identified as the superconducting gap.

Outside the gap we observe three distinct satellite features labeled A, B and C in Fig. 3b, that originate from the coupling of electrons to collective excitations (Bosonic modes such as spin excitations or phonons) in the sample. The signature of a mode in STS spectra can arise from two processes. The first is a direct modification of the electronic density of states where the coupling of electrons to a Bosonic mode renormalizes the self-energy. The second process occurs when the tunneling electron excites a boson before the transition to its final state (inelastic tunneling). The analysis required to determine the mode energy $E$ depends on which channel dominates the spectrum. Many details, including the dispersion of the bosonic mode and the relative strength of the two processes, determine the lineshape of the resultant feature in the spectrum. Both phonons and spin-excitations have been observed by STS and in most instances (like phonons in graphene[10] or spin excitations in magnetic clusters[11]) the inelastic channel is stronger. This dictates our analysis for the mode energy. We note that while the existence of these modes on both sides of the Fermi energy is clear, in the 21 K sample we do observe that there are sometimes 1 to 2 meV differences between positive and negative sides. This is possibly due to a background ldos asymmetry, perhaps due to the underlying band structure such as a van Hove singularity that makes the extraction of the mode energy less precise on the negative bias side.

STM of cleaved PLCCO shows atomic scale features but not the periodic lattice. This makes it difficult to gauge the quality of the tip simply based on STM images of

PLCCO surfaces. We therefore adapted a two-step process to study PLCCO samples. While this procedure makes the experiment more difficult, it provides us with a greater control over the quality of the tip. Our process involves first characterizing the tip on the prototypical high temperature superconductor $Bi_2Sr_2CaCu_2O_8$ (Bi2212) which has a well known surface (Fig. SI7a) and spectra (Fig. SI7b). Once a tip shows atomic resolution and stable spectra on Bi2212, we proceed with PLCCO (Fig. SI7c and d).

We used a few different techniques to ensure the reliability of our data. First, we characterized the tips on Bi2212 prior to using it on PLCCO as discussed in the previous paragraph. Although STM is the ultimate tool to study topography at the nanoscale, we also characterized cleaved samples by scanning electron microscopy (SEM). SEM images of cleaved PLCCO samples are shown in Fig. SI8a and b. The images show that there are indeed large micron sized regions, which are flat and potentially suitable for STM studies.

STM images reveal that there are steps (Fig. SI8c) and flat regions (corrugation of less than 100 pm, (0.1 Å) over 128 Å and less than 0.5 Å over larger areas (256 Å) as shown in the inset to Fig. SI8e and f. These regions were accessed by walking to different spots using the coarse walking mechanism of the STM, which proved critical for studies on PLCCO. The scale of the contrast in the images is small and suggests density of states effects rather than surface roughness. As a comparison, single atoms on the surface are typically 0.5 Å -1 Å high, while steps are heights are larger (of the order of a few Å to tens of Å depending on the sample). For our study we chose spots on the basis of stable spectra with a superconducting gap. Disordered regions on the sample, which showed semiconducting spectra, were not studied for this work.

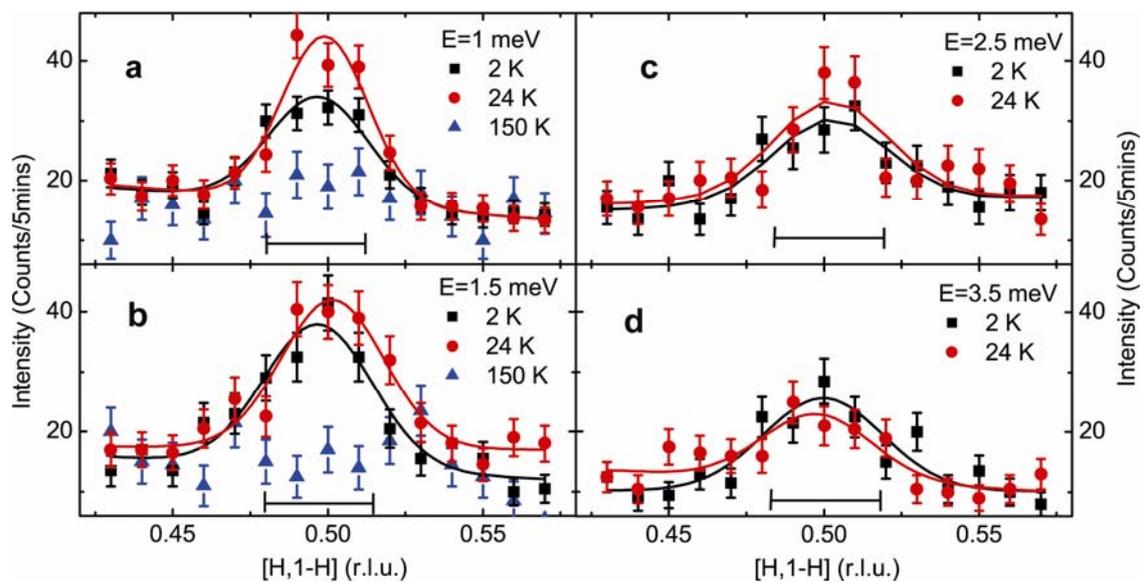

**Figure SI1 Constant-energy scans using cold triple-axis SPINS on the 24 K PLCCO.** Transverse $Q$-scans through $Q = (0.5, 0.5, 0)$ for (a) $E = 1$ meV; (b) $E = 1.5$ meV; (c) $E = 2.5$ meV; (d) $E = 3.5$ meV at different temperatures; the horizontal bars are the instrument resolutions; the solid lines are Gaussian fits on linear backgrounds. The raw

data at 150 K in Figs. SI1a and SI1c have been down shifted by 9 counts and 12 counts, respectively.

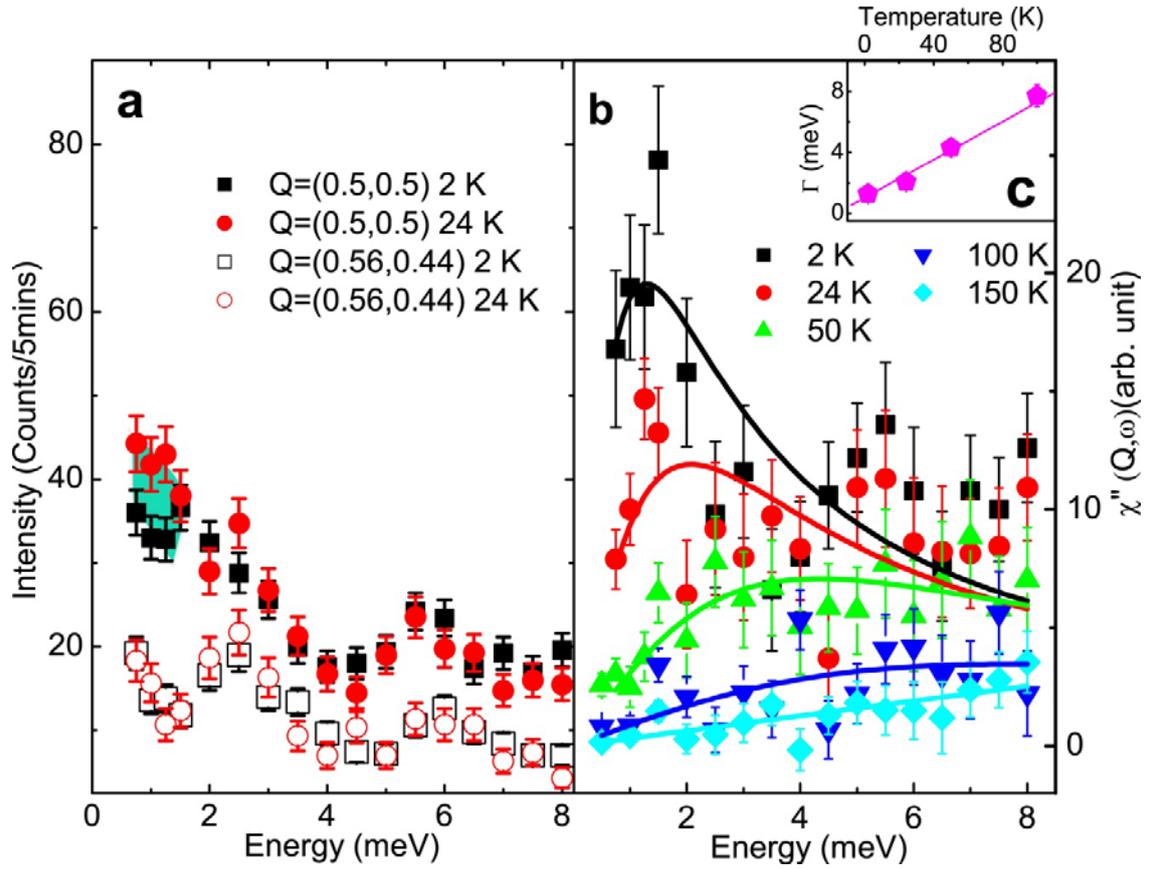

**Figure SI2 Energy dependence of the scattering at $Q = (0.5, 0.5, 0)$ rlu and $\chi''(\omega)$ for the 24 K PLCCO sample.** (a) Energy scans at the AF zone center [$Q = (0.5,0.5,0)$] and background [$Q = (0.56,0.44,0)$] positions at 2 K and 24 K. We did not plot similar data at other temperatures. A spin pseudo gap opens up below $T_c$ in $S(Q,\omega)$; the corresponding $\chi''(Q,\omega)$ are plotted in (b). The solid lines in (b) are fits to $x''(\omega) \propto \Gamma\omega/(\Gamma^2 + \omega^2)$, and (c) The temperature dependence of the relaxation rate $\Gamma$ at different temperatures.

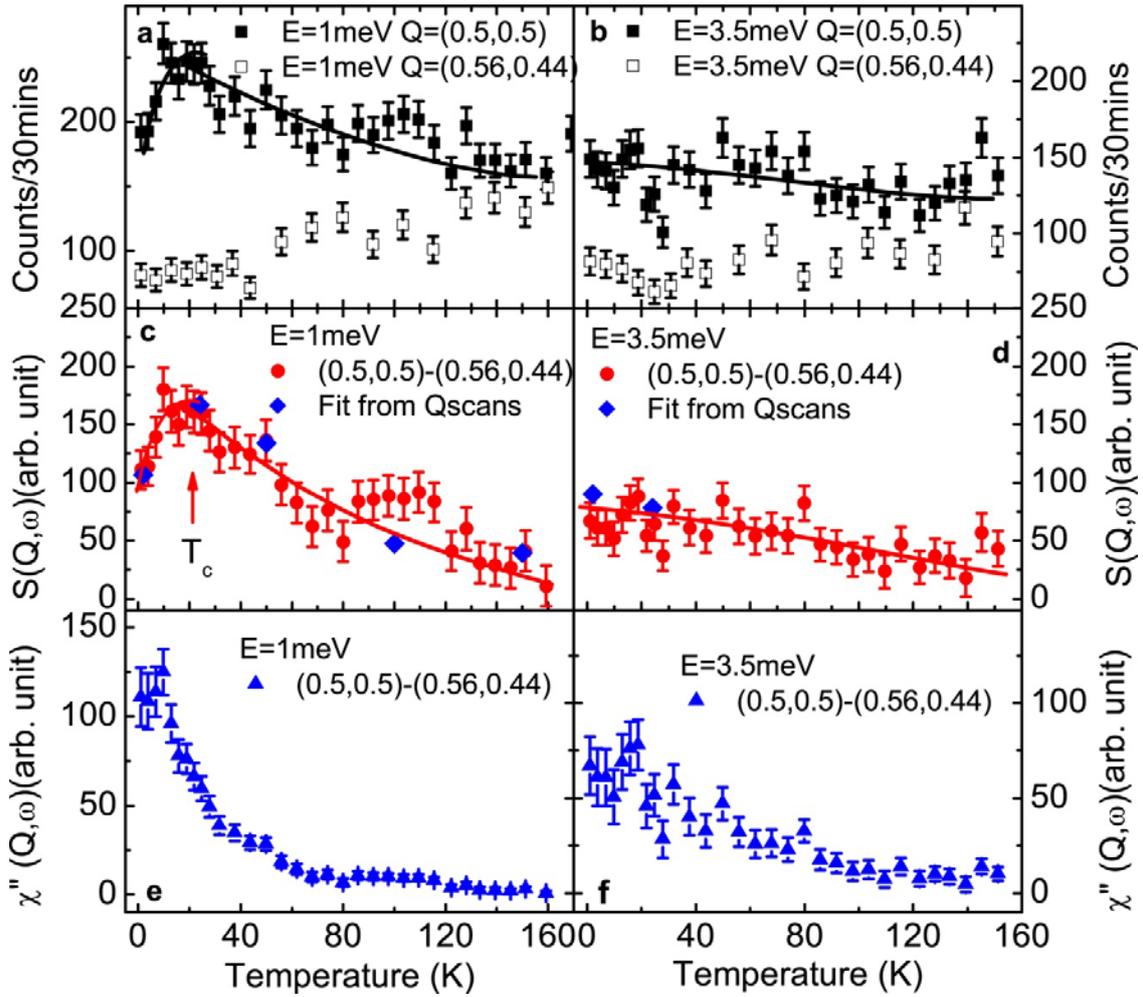

**Figure SI3 Temperature dependence of the scattering at 1 meV and 3.5 meV near AF ordering wave vector and background positions.** (a), (b) Temperature dependence of the scattering at the signal [$Q = (0.5,0.5,0)$] and background [$Q = (0.56,0.44,0)$] positions for $E =1$ meV and $E = 3.5$ meV, respectively. (c) The background subtracted magnetic scattering at $E = 1$ meV. (d) The background subtracted magnetic scattering at

$E = 3.5$ meV showing no anomaly across $T_c$; the solid lines are guided to the eye. (e) and (f) plot the corresponding $\chi''(Q,\omega)$ at 1 meV and 3.5 meV, respectively.

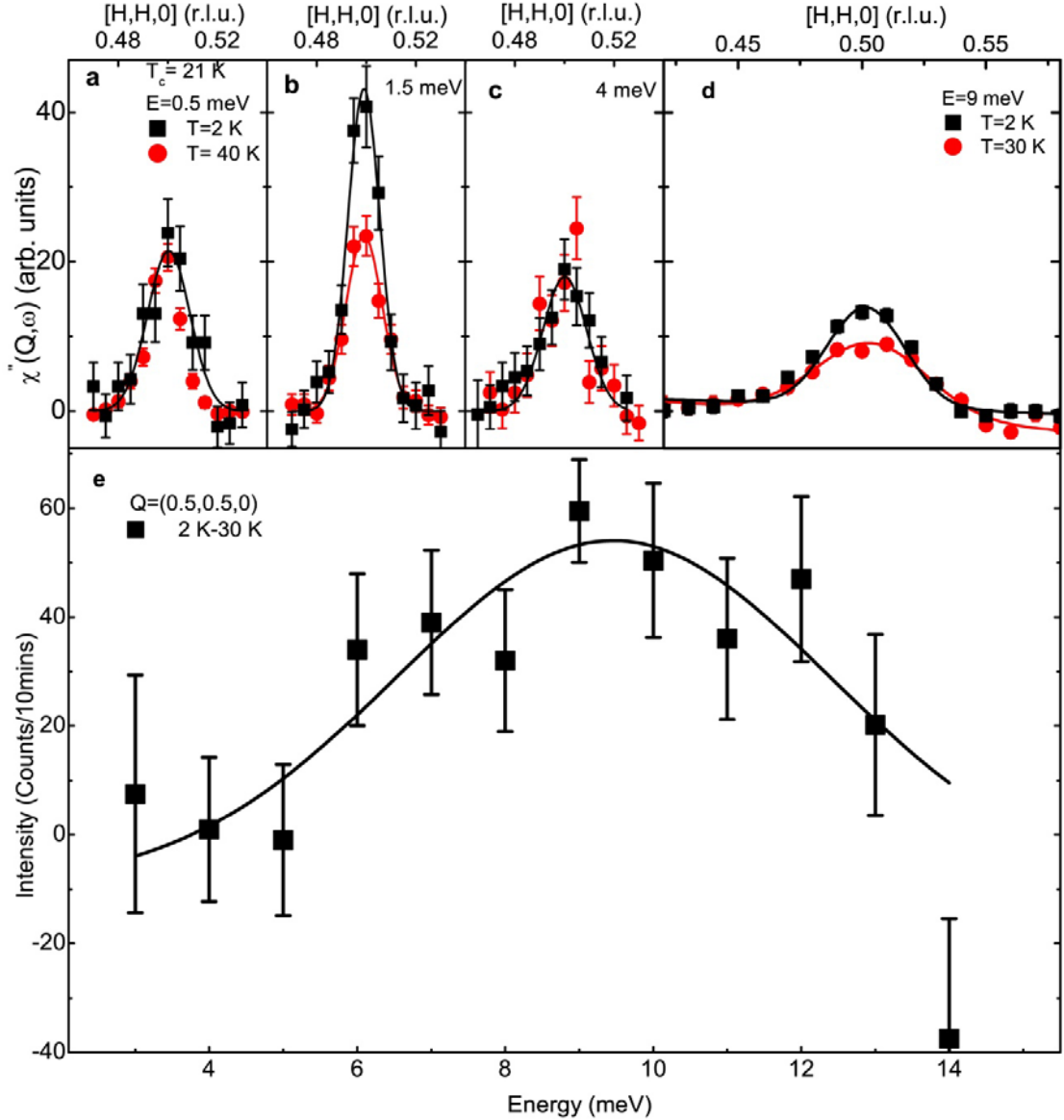

**Figure SI4 Temperature dependence of the $\chi''(Q,\omega)$ at 0.5, 1.5, 4 and 9 meV near AF ordering wave vector.** (a-c) Data are from Ref. 2 taken on SPINS; (d) Data from Ref. 3 taken on a thermal triple-axis; (e) Temperature dependence plot for constant-$Q$ scans below and above $T_c$ at $Q = (0.5,0.5,0)$. Data in (d,e) are from Ref. 3.

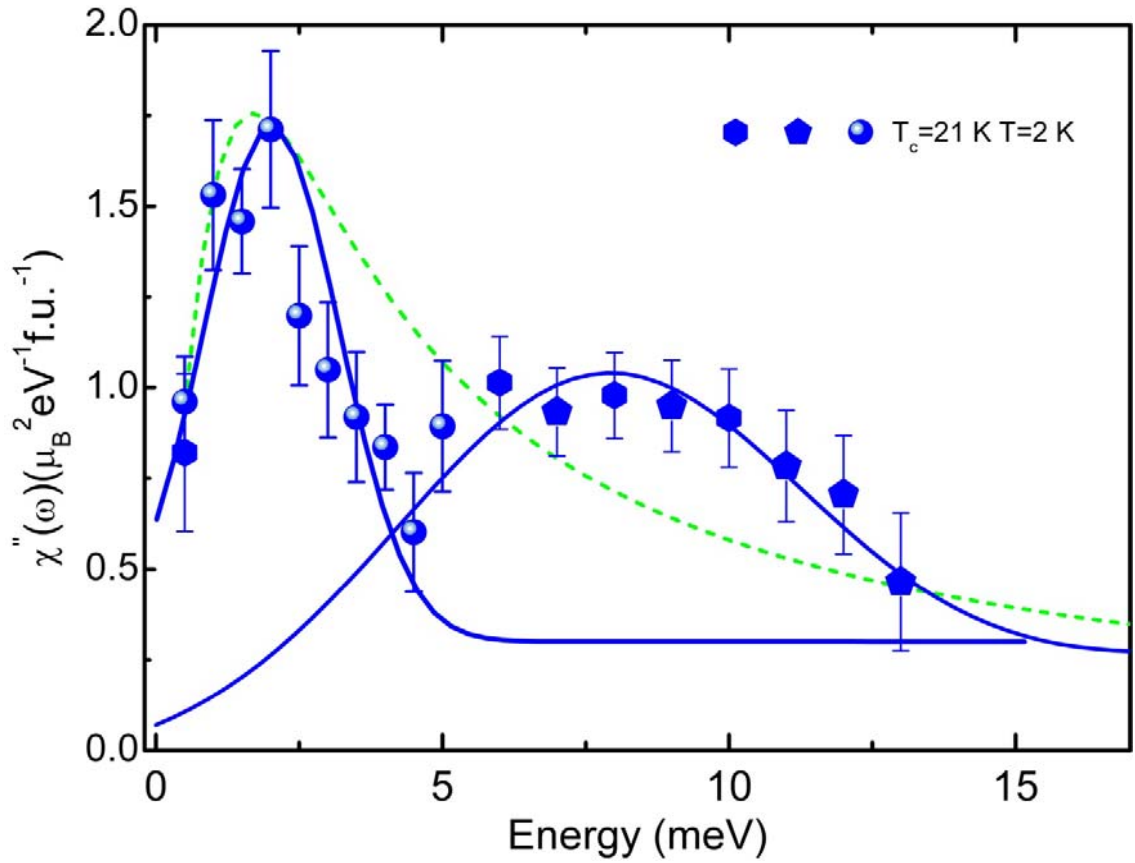

**Figure SI5 Combined cold and thermal triple-axis spectrometer data normalized to time-of-flight data in absolute units.** Two well defined modes can be observed near 2 meV and 9 meV. The green dashed line shows an attempted fit using a modified Lorentizian which clearly cannot fit the data. The solid lines are guides-to-the-eye based on weighted Gaussian fit. The presence of two modes is also confirmed by constant-energy scans shown in Fig. SI4 a-d.

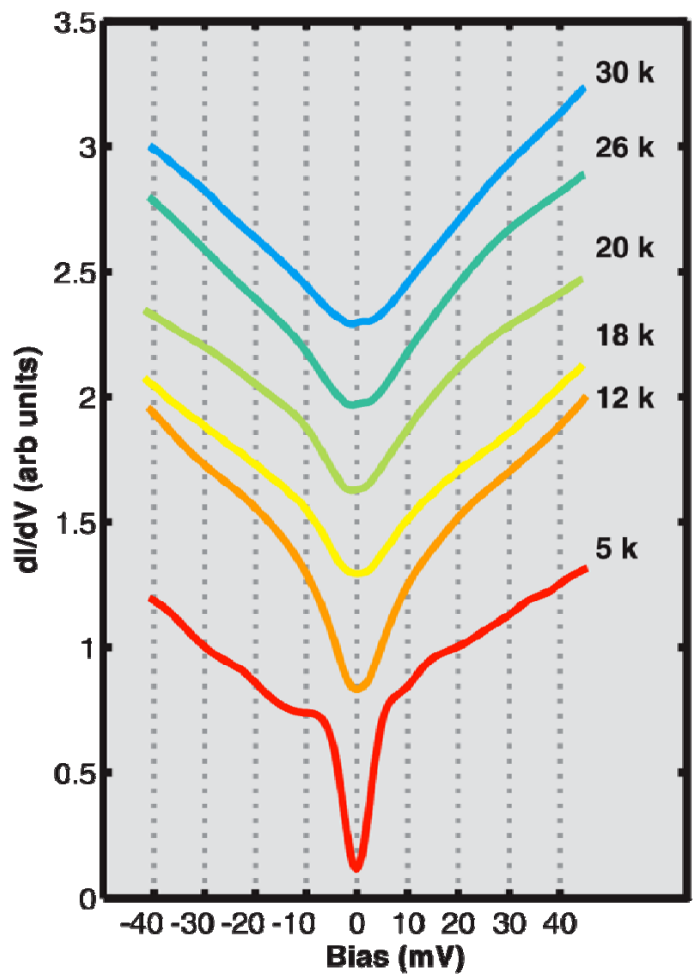

**Figure SI6 Temperature dependence of averaged dI/dV spectra obtained on the 21 K PLCCO**. The data show mute coherence peaks and vanishing superconducting gap above the superconducting transition temperature. The spectra are offset for clarity.

# Tip characterization

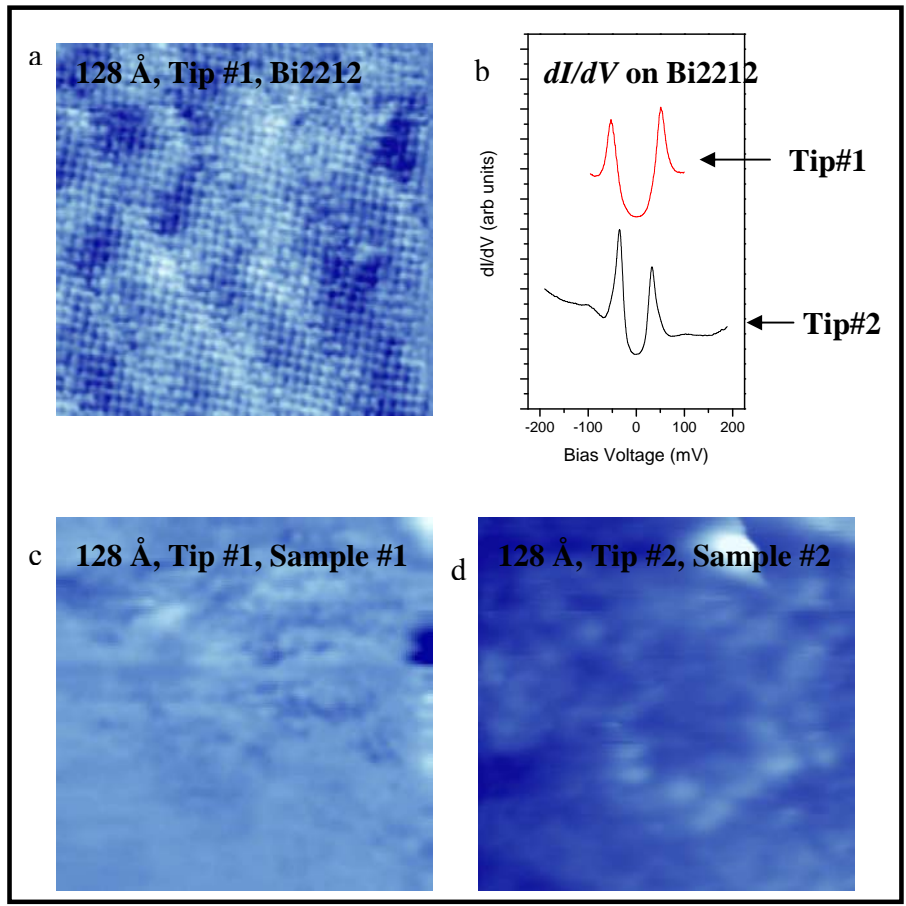

**Figure SI7 Tip characterization on Bi2212 prior to studying PLCCO.** a) STM image of Bi2212 with tip#1 showing atomic resolution, which was then used to study sample #1 (24K PLCCO) shown in c. b) dI/dV spectra of Bi2212 with two different tips used to study 21K and 24K samples respectively. The tip is deemed stable if it shows atoms and super-lattice modulation on Bi2212 and the spectra on Bi2212 clearly show sharp coherence peaks as well as the phonon modes at higher energies outside the gap. c) and d) PLCCO images with stable tips, tip #1 and tip #2.

# Characterization of cleaved samples

## SEM Images of cleaved PLCCO

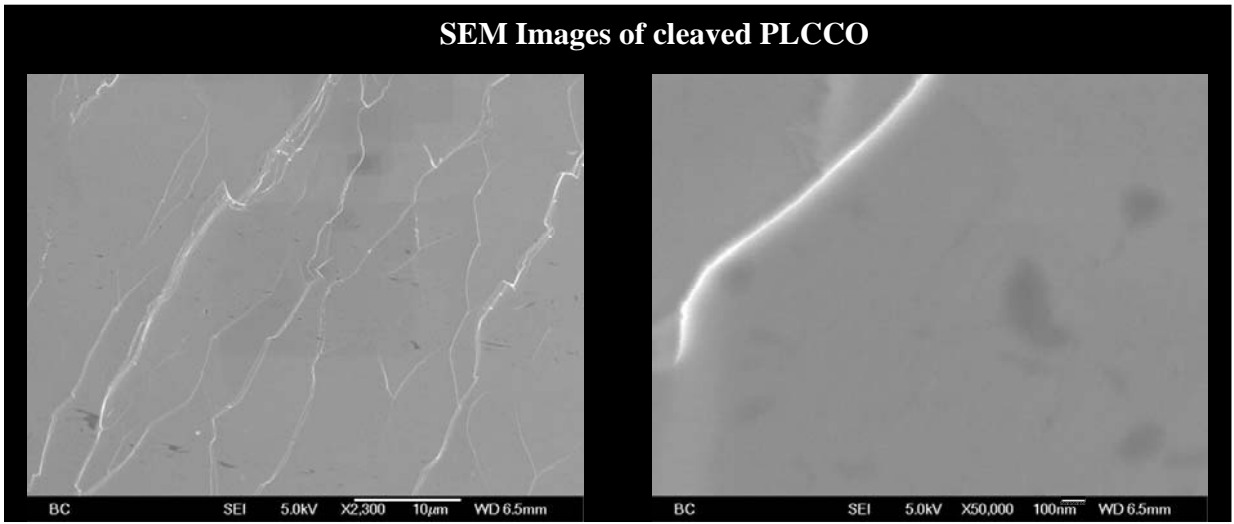

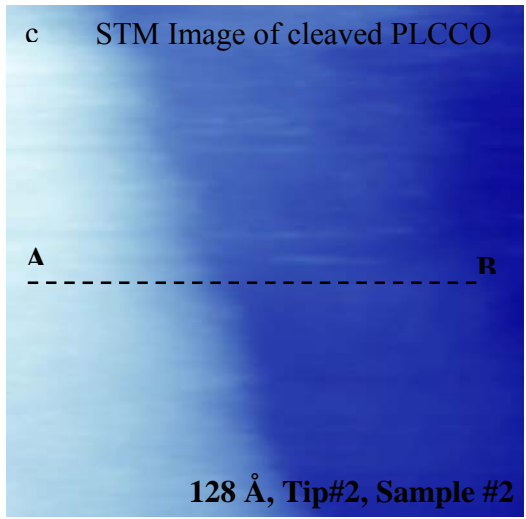

c  STM Image of cleaved PLCCO

A ----------------- B

128 Å, Tip#2, Sample #2

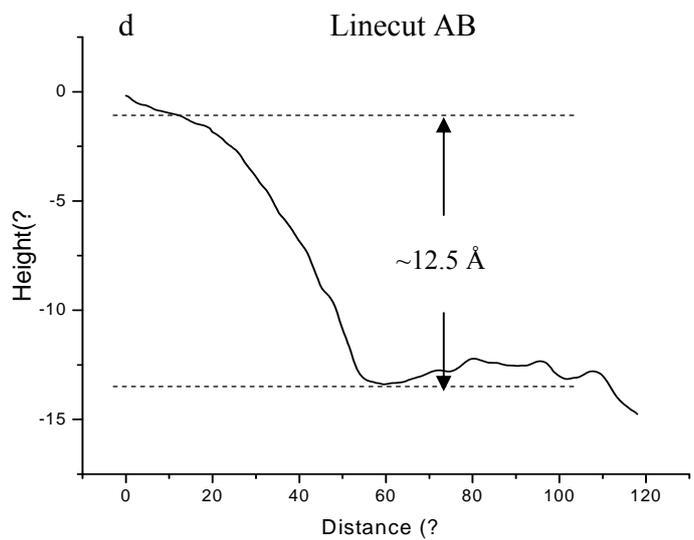

d  Linecut AB

~12.5 Å

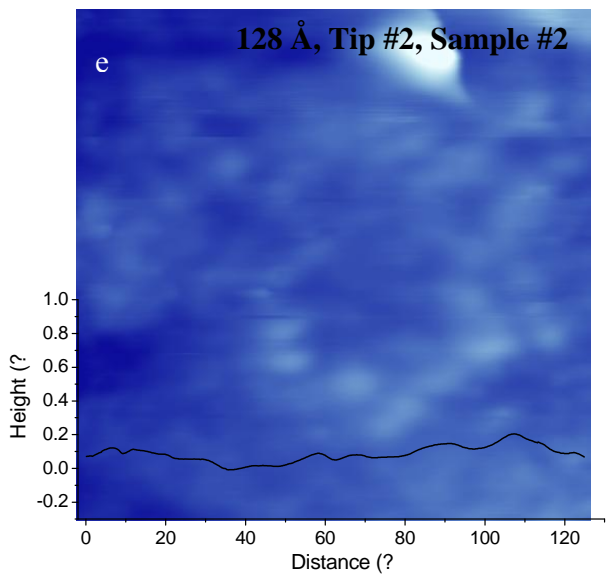

128 Å, Tip #2, Sample #2

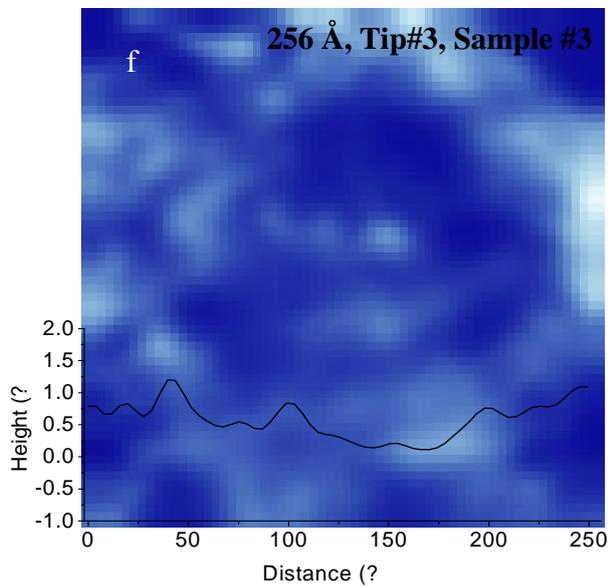

256 Å, Tip#3, Sample #3

**Figure SI8 PLCCO surface characterization with SEM and STM** a) and b) SEM images of PLCCO clearly showing micron sized regions with flat terraces and steps. c) STM image of step on PLCCO. The associated linecut shown in d) reveals that the step height is ~12.5Å consistent with the c-axis height. While this is not sufficient information to identify the cleave plane, it tells us that there are regular terraces and steps on PLCCO. e) and f) STM images of 21K PLCCO with well characterized tips. The images show flat regions (corrugation of less than 100 pm, (0.1 Å) over 128 Å and less than 0.5 Å over larger areas (256 Å) as shown in the inset. Most features of this size can be attributed to density of states effects. Note that the sharpness of the tip is also revealed by the fact that we can resolve atomic scale features on which are a few Å in width.